# Probing General Relativity-Induced Decoherence Using an on-chip Sagnac Interferometer

MOHAMED ELKABBASH[1]

[1]*James C. Wyant College of Optical Sciences, University of Arizona, Tucson, Arizona 85721, USA*
*melkabbash@arizona.edu*

**The intersection of quantum mechanics and general relativity remains an open frontier in fundamental physics, with few experimentally accessible phenomena connecting the two. Recent theoretical proposals suggest that relativistic proper time can act as a source of decoherence in quantum systems, providing a testable overlap between the two theories. Here, we propose a chip-integrated Sagnac interferometer where rotation induces a proper time difference between clockwise and counterclockwise single-photon paths. When this time delay exceeds the photon's coherence time, interference visibility is predicted to decrease, offering a direct signature of relativistic time dilation-induced decoherence. We theoretically derive the proper time difference arising from the Sagnac effect and estimate that for a loop radius of 18.9 cm and a rotation speed of 1000 rad/s, decoherence should occur for single-photon wavepackets with a coherence time of 10 femtoseconds. We also present a practical chip design that accommodates the required high-speed mechanical rotation and includes an all-optical readout scheme to eliminate wiring constraints. This approach enables a stable, on-chip implementation using realistic parameters, with rotation speed serving as a continuously tunable knob to control decoherence. Our platform opens a new route for experimentally probing the interplay between quantum coherence and relativistic proper time in a scalable and compact form.**

## 1. INTRODUCTION

The interplay between quantum mechanics and general relativity remains one of the most profound open questions in modern physics. While both theories have been extensively validated within their respective domains—quantum mechanics at microscopic scales and general relativity at macroscopic scales—their direct experimental overlap has been limited. Recently, several experiments have been proposed where effects predicted by one framework affect the outcomes predicted by the other framework.

The first experimentally meaningful quantity that connects these two frameworks is entanglement. Recent proposals suggest that if gravity is a quantum entity, it should be capable of mediating entanglement between two masses. Bose et al. [1] proposed an experiment where two masses are prepared in a spatial superposition where the positions are entangled with their spin degree of freedom. If gravity is quantum, it can mediate entanglement between the masses by virtue of their position superpositions. This entanglement is then transferred to the spins, which act as entanglement witnesses.

Beyond entanglement, another experimentally testable bridge between general relativity and quantum mechanics is proper time. Proper time, in general relativity, is the time measured by a clock moving along a specific path (or worldline) in curved spacetime. It is the invariant interval between two events as experienced by an observer who passes through both events. Zych et al. [2] proposed that interferometric visibility loss can serve as a quantum test of proper time. Their work showed that a quantum system with internal energy states acting as a clock experiences gravitational time dilation, leading to a difference in the proper time between the two paths. This difference leads to distinguishability because it effectively allows an observer (even in principle) to extract which-path information from the system's internal clock. This distinguishability results in the loss of quantum coherence, often interpreted as a partial or complete collapse of the wavefunction in the interferometric setup.

This proposal was later extended to quantum optical systems, where single photons with finite coherence times experience similar decoherence effects due to gravitational time dilation [3]. In this scenario, the photon's coherence time plays the role of an internal clock, and gravitational time dilation causes a loss of interferometric visibility when the time delay exceeds the photon's coherence time. The proposed experiment employed a Mach-Zehnder interferometer (MZI) to test the influence of gravitational time dilation on single-photon interference. In this setup, a single photon is split into two spatially separated paths at different heights within Earth's gravitational field. Since each path corresponds to a different gravitational potential, the photon wavepacket experiences a difference in arrival times due to gravitational time dilation. When the time delay between the two arms becomes comparable to or exceeds the photon's coherence time, which-path information becomes, in principle, accessible—leading to a measurable reduction in interference visibility. This loss of coherence directly probes general relativistic time dilation as a mechanism for quantum decoherence.

While the MZI-based approach provides a conceptual framework for testing gravitational time dilation-induced decoherence, its practical realization faces significant challenges. Achieving a sufficient difference in gravitational potential between the interferometer arms requires a vertical height separation of several meters, which is difficult to implement in a typical lab. To accumulate a measurable time delay, long optical fibers—kilometers in length—must be used in each arm. Although using ultrashort single-photon pulses can reduce the required time delay, the associated broad spectral bandwidth leads to group velocity dispersion over long propagation distances, causing wavepacket broadening. Even if dispersion is compensated using pre-chirping to recompress the wavepackets at the second beamsplitter, photon losses become increasingly problematic due to fiber attenuation, especially for broadband photons, leading to very low detection rates. Furthermore, a proper control experiment would require rotating the entire setup to verify that visibility loss arises from gravitational time dilation and not other decoherence sources. Such mechanical rotation introduces alignment errors and other instabilities, making it difficult to isolate the relativistic effect. These constraints significantly limit the feasibility of the MZI-based approach in a controlled tabletop experiment.

To overcome these limitations, here we propose an alternative approach based on a Sagnac interferometer, which naturally introduces a difference in proper time between clockwise (CW) and counterclockwise (CCW) propagating beams due to rotation. In our scheme, a single photon is split into a superposition of CW and CCW paths within an integrated Sagnac loop. When the platform is rotated, relativistic time dilation causes a measurable difference in the proper time experienced by each path. If this difference exceeds the photon's coherence time, the distinguishability between the two paths increases, resulting in a measurable drop in interference visibility. By monitoring visibility as a function of rotation speed, we can directly observe the transition from quantum coherence to decoherence arising from proper time differences, without relying on gravitational potentials or extended optical paths.

This approach offers several advantages over the MZI-gravitational time dilation scheme. First, it enables an entirely on-chip implementation using realistic experimental parameters, significantly reducing the complexity and instability associated with bulk optical setups. Unlike MZI interferometers, the Sagnac interferometer is a shared-path geometry, inherently more stable and less prone to phase noise and misalignment. Moreover, the integrated nature of the platform ensures minimal optical losses and allows for seamless coupling to single-photon sources and detectors. Crucially, the use of rotation speed as a continuously tunable parameter provides a direct and controllable knob to adjust the proper time difference between the CW and CCW paths. This allows for systematic exploration of the resulting visibility loss and offers a clear experimental signature of relativistic time dilation-induced decoherence, without requiring extreme geometries or discrete changes in gravitational potential.

## 2. THEORETICAL ANALYSIS

Here, we explore the general relativistic-induced decoherence in single photons through time dilation due to rotation, i.e., the Sagnac effect. The Sagnac effect introduces a time dilation between the CW and CCW propagating beams. Previous experiments have demonstrated the Sagnac effect with single photons[3], and it is widely integrated on photonic chips[4]. The phase shift due to rotation is given by: $\Delta\phi = (8\pi\Omega A)/(c\lambda)$, where $\Omega, A, c$, and $\lambda$, are the rotation speed in rad/s, the area enclosed by the Sagnac loop, the speed of light, and the optical wavelength, respectively. This phase shift corresponds to a leading to a time delay given by $\Delta t = (\Delta\phi\lambda)/(2\pi c) = (4\Omega A)/(c^2)$.

The interference visibility, determined by the coherence between the two time-evolved quantum states, is given by: $V = |\langle \tau_{CW} | \tau_{CCW} \rangle|$, which depends on the distinguishability $D$ of the two paths, satisfying the complementarity relation: $V^2 + D^2 = 1$. Using the time delay expression from the Sagnac effect, the visibility follows an exponential decay:

$$V = e^{-(\Delta t/t_c)^2} = e^{-(4\Omega A/(c^2 t_c))^2} \quad (1)$$

Where $t_c$ is the coherence time of the single photon which depends on its spectral bandwidth.

*Feasibility analysis:*
While it may seem advantageous to use a high-Q ring resonator cavity in a Sagnac interferometer due to its enhanced finesse this approach is not feasible. The issue arises because ring resonators act as both *spectral and temporal* filters, which increases the temporal spread (Δt) of the photon wavepacket as the quality factor Q increases. According to the time–frequency uncertainty principle, $\Delta t \approx Q/(4\pi f)$, so for visible or near-infrared photons ($f \approx 10^{14}\,Hz$), even moderate $Q$-factors result in significant time spreads. Meanwhile, the Sagnac-induced time delay in a ring resonator is given by $\Delta t \approx (\lambda_0 \cdot \Omega \cdot Q \cdot r)/c^2$, where $\Omega$ is the rotation speed and $r$ is the resonator radius which arises from the multiple trips the CW and CCW single photon states take inside the cavity. For the Sagnac time delay to match the increased coherence time (and thus allow visibility degradation), this relationship demands an unphysically large ring radius on the order of $10^4$ meters. Therefore, rather than making decoherence effects more observable, the use of a ring resonator suppresses them unless extremely large and impractical device dimensions are used. This makes the ring resonator approach unsuitable for demonstrating proper-time-induced decoherence in a realistic, chip-based Sagnac experiment.

On the other hand, for a simple Sagnac loop, similar to the one shown in FIG.1, with a radius of 18.9 cm and a rotation speed of 1000 rad/s, we expect a time dilation exceeding the coherence time of a single photon wavepacket with $t_c = 10\,fs$. Generating single photons with this coherence time is possible, for example, using light-field synthesis followed by spontaneous parametric down conversion[5]. Moreover, rotating a chip by 1000 rad/s is technically possible. Finally, single-photon Sagnac interference has already been experimentally observed using fiber-based Sagnac loops[3].

## 3. PROPOSED EXPERIMENT

We propose an experimental setup, taking into account practical design constrains, that would enable the demonstration of the time-dilation decoherence using on-chip Sagnac interferometers [4]. The experimental setup is illustrated in FIG. 1. A laser pulse generates a single-photon pair via spontaneous parametric down-conversion. One photon from the pair is detected to herald the presence of the other, which is directed to the integrated photonic chip. The photon is coupled into the chip through a grating coupler positioned at the center of the device such that the rotation of the chip does not lead to a change in the grating position. A polarization independent grating coupler could be necessary. Alternatively, a half-wave plate in sync with the chip can rotate to ensure that the incident photon polarization is maintained relative to the grating coupler. After coupling, the photon enters a directional coupler, which splits it into a superposition of two paths (CW and CCW) within the loop structure forming the Sagnac interferometer.

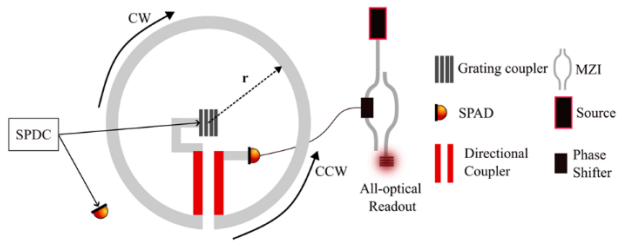

Fig. 1. Schematic of the experimental setup required for testing the overlap between General Relativistic effects and Quantum Mechanics. The photonic chip contains an on-chip Sagnac loop with a grating coupler in the center of the chip. The rotation induces time dilation between clockwise (CW) and counter-clockwise (CCW) photons. If the time dilation exceeds the coherence time of the single photon wavepacket, the interference visibility will decrease. The visibility is determined through the amplitude of the sinusoidal dependence of the photon counts on the rotation speed which is measured through an integrated single photon detector, here a SPAD. The entire chip sits on a chuck that rotates up to 1000 rad/s. Due to the required high-speed rotations, wires used for readout should be eliminated while other on-chip electronics will be powered by a battery that can be physically attached to the chuck. The readout is done through modifying the output from an MZI using a phase-shifter controlled by the SPAD. For this experiment, an on-chip heterogeneously integrated laser source is used to avoid difficulty in coupling light to a rotating chip. Outcoupled light from the grating indicates a photon detection event.

Due to the rotation-induced time dilation, the relative time delay between the CW and CCW components affects their interference. This results in a modified probability for the photon to be either transmitted or reflected at the directional coupler. A single-photon detector such as a single photon avalanche diode (SPAD) detects the transmitted photons, and their count is recorded as a function of the rotation speed.

As the rotation speed increases, the interference visibility decreases, signaling the transition from quantum coherence to decoherence. Initially, the transmitted and reflected photon counts vary with the phase shift due to interference. As decoherence sets in, this phase dependence weakens, and the visibility gradually drops. Beyond a critical rotation speed, the interference pattern disappears entirely, and the detection probabilities settle at 50/50, assuming the directional coupler splits the power equally, indicating that the photon is no longer in a coherent superposition but instead behaves as a classical mixture.

***On-chip all-optical readout:*** Since the photonic chip is rotating and to avoid wiring the chip to perform the readout, we propose here an all-optical readout scheme.

A laser is integrated on-chip and its light is directly coupled to an MZI with a phase shifter such that one of the MZI waveguides is terminated with a grating coupler while the other is terminated with a beam dump. Initially, the shifter's phase is tuned to ensure that no light is outcoupled from the grating. For signal readout, the SPAD detection signal can be used to change the phase shifter settings. In the absence of a detection event, the interference at the MZI's output ports prevents light from coupling out. When a photon is detected, the SPAD generates an electrical signal that induces a phase shift in the MZI, redistributing light and directing it toward the output grating coupler. This all-optical readout mechanism eliminates the need for electrical wiring, reducing the mechanical complexity when the chip is rotating**.**

## 4. CONCLUSION

In this work, we proposed and analyzed an experimental scheme to probe general relativistic time dilation-induced decoherence using an on-chip Sagnac interferometer. The proper time difference introduced by rotation between clockwise and counterclockwise single-photon paths enables direct measurement of visibility loss as a function of rotation speed. We showed that, unlike previous proposals based on MZIs and gravitational potential differences, our scheme can be implemented using realistic parameters entirely within an integrated photonic platform. This on-chip geometry offers enhanced stability due to its shared-path configuration, eliminates long propagation paths that introduce photon loss and dispersion, and provides a continuously tunable control parameter which is the chip's the rotation speed. Successful implementation of this proposal would be an important step toward bridging quantum mechanics and general relativity by enabling controlled and observable overlap between these two foundational frameworks. Such capability opens new directions in relativistic quantum information, quantum sensing, and fundamental tests of physics using compact, table-top platforms.

**Funding.** The author received no specific funding for this work.

**Acknowledgment.** The author acknowledges the fruitful discussions with Dr. Ahmed Roman, Dr. Asem Habashi, and Dr. Islam Shalaby. The author would also like to acknowledge Mr. Abrar Fahim Liaf for help with formatting the manuscript.

**Disclosures**. The authors declare no conflicts of interest.

**Data Availability Statement (DAS).** Data are available upon request from the author.